\documentclass[aps,twocolumn,showpacs,groupedaddress]{revtex4}
\usepackage{graphicx}

\begin{document}

\title{Geometric phase in open systems: beyond the Markov approximation and
weak coupling limit}
\author{X. X. Yi$^{1,2}$,  D. M. Tong$^2$, L. C. Wang$^1$,
L. C. Kwek$^3$, C. H. Oh$^2$}
\affiliation{$^1$Department of Physics, Dalian University of
Technology, Dalian 116024, China\\
$^2$Department of Physics, National University of  Singapore, 10
Kent Ridge Crescent, Singapore 119260\\
$^3$ Department of Natural Sciences, National Institute of
Education, Nanyang Technological University, 1 Nanyang Walk,
Singapore 637616}

\date{\today}

\begin{abstract}
Beyond the quantum Markov approximation and the weak coupling
limit, we present a general theory to calculate the geometric
phase for open systems with and without conserved energy. As an
example, the geometric phase for a two-level system coupling both
dephasingly and dissipatively to its environment is calculated.
Comparison with the results from quantum trajectory analysis is
presented and discussed.
\end{abstract}

 \maketitle
\section{introduction}
Since Berry's discovery \cite{berry84} that a state of a quantum
system can acquire a phase of purely geometric origin when the
Hamiltonian of the system undergoes a cyclic and  adiabatic
change, there have been numerous proposals for generalizations,
including the geometric phase for nonadiabatic, noncyclic, and
nonunitary evolution \cite{shapere89}, the geometric phase for
mixed states \cite{uhlmann86,sjoqvist00a}, the geometric phase in
systems with quantum field driving and vacuum induced effects
\cite{fuentes02}, as well as the geometric phase in coupled
bipartite systems\cite{yi04}.

Recently, the attention has turned to studying geometric phase in
open systems, this is motivated by the fact that all realistic
system are coupled, at least weakly, to their environment. From
the perspective of application, the use of geometric phases in the
implementation of fault-tolerant quantum gates
\cite{zanardi99,jones99,ekert00,falci00} requires  the study of
geometric phases in more realistic systems. For instance, the
system which carries information may decoher from a quantum
superposition into statistical mixtures, and this effect, called
decoherence, is the most important limiting factor for quantum
computing.

The study of geometric phase in open systems may be traced back to
1980s, when Garrison   and Wright \cite{garrison88}   first
touched  on this issue by describing open system evolution in
terms of  non-Hermitian Hamiltonian. This is a pure state
analysis, so it did not address the problem of geometric phases
for mixed states. Toward the geometric phase for mixed states in
open systems, the approaches used include solving the master
equation of the system
\cite{fonsera02,aguiera03,ellinas89,gamliel89,kamleitner04},
employing a quantum trajectory analysis \cite{nazir02,carollo03}
or Krauss operators \cite{marzlin04}, and the perturbative
expansions \cite{whitney03,gaitan98}. Some interesting results
were achieved that may be  briefly summarized as follows:
nonhermitian Hamiltonian lead to a modification of Berry's phase
\cite{garrison88, whitney03}, stochastically evolving magnetic
fields produce both energy shift and broadening \cite{gaitan98},
phenomenological weakly dissipative Liouvillians alter Berry's
phase by introducing an imaginary correction \cite{ellinas89} or
lead to  damping and mixing of the density matrix elements
\cite{gamliel89}. However, almost all these studies are performed
for dissipative systems under various approximations,  thus the
representations are approximately  applicable for systems whose
energy is not conserved. Quantum trajectory analysis
\cite{nazir02,carollo03} that uses the quantum jump approach is
available for open systems with conserved energy. Its starting
point, however, is the master equation, a result within the
quantum Markov approximation and in the weak coupling limit.
Beyond the quantum Markov approximation and the weak coupling
limit, the geometric phase of a two-level system {\it with quantum
field driving} was analyzed \cite{yi05}, where the whole system
(the two-level system plus the quantum field) was assumed subject
to dephasing, the subsystem(two-level system in that paper) may
still acquire geometric phase even when the whole system in its
pointer states. This is an ideal situation to show the vacuum
effects on the geometric phase of the subsystem, as well as the
decoherence effects on the geometric phases regardless of its
feasibility of experimental realization. However, beyond the
Markov approximation and the weak coupling limit, the geometric
phase for a dissipative system remains untouched. In this paper,
we will deal with the geometric phase in open systems, beyond the
Markov approximation and weak coupling limit.

The structure of this paper is organized as follows. In Sec. {\rm
II} the exact solution and calculation of the geometric phase of a
system dephasingly coupled to its environment is presented, an
example to detail the representation and a discussion on
physical realization are given in Sec. {\rm III}. In Sec. {\rm
IV}, we present an example to show the calculation of geometric
phases in dissipative systems. Finally we conclude in Sec.{\rm V}.

\section{geometric phase in dephasing systems: general
formulation}
 In this section, we investigate the behavior of
geometric phase of a quantum system subject to decoherence. In
order to make a comparison with the results from the quantum jump
approach, we consider the quantum system without any field driving
except the environment. So, it is not directly relevant to our
previous study\cite{yi05}. The environment that leads to
decoherence may originate form the vacuum fluctuations or the
background radiations. We restrict ourselves here to consider the
case where the system-environment coupling $H_I$ and the free
system Hamiltonian $H_S$ commute. This is the situation of
dephasing and the exact analytical dynamics  may be obtained. On
the other hand, the evolution of  a system with such properties
may be described by the master equation when the Markov
approximation and the weak coupling assumption apply,  this kind
of decoherence would not change the geometric phase of the quantum
system by the quantum jump approach\cite{nazir02,carollo03}.
However, as you will see, this is not the case from the viewpoint
of interferometry considered in this paper.

We consider a situation described by a Hamiltonian of the form
\begin{equation}
H=H_S+H_B+H_I,
\end{equation}
where $H_S$ describes the free Hamiltonian of the system, $H_B$
stands for the Hamiltonian of the environment, and $H_I$
represents the system-environment couplings. The environment and
the system Hamiltonian may be taken arbitrary but with constraints
$[H_S,H_I]=0$. Let us suppose that the interaction Hamiltonian
$H_I$ has the form (setting $\hbar=1$)
\begin{equation}
H_I=\sum_m X_m(\Gamma_m^{\dagger}+\Gamma_m),\label{gm}
\end{equation}
where the $X_m, (m=1,2,...,M)$ are the system operators satisfying
$[H_S, X_m]=0,$ and the $\Gamma_m$ represent  environment
operators that may  take any form in general.  Commutation
relation $[H_S, X_m]=0$ enables us to write the time evolution
operator for the whole systems(system+environment) as
\begin{equation}
U(t)=e^{-iHt}=e^{-iH_St}\sum_{m}U_{m}(t)|E_{m}\rangle\langle
E_{m}|
\end{equation}
with $U_{m}(t)$, a function of environment operators satisfying
\begin{eqnarray}
i\frac{\partial}{\partial
t}U_{m}(t)&=&H_{e,m}U_{m}(t),\nonumber\\
H_{e,m}&=&H_B+\sum_n e_{n}^m(\Gamma_n+\Gamma_n^{\dagger}).
\end{eqnarray}
Here, $|E_m\rangle$ stands for the eigenstate of $H_S$ with
eigenvalue $E_m$\cite{note1}, while $e_n^m$ denotes the eigenvalue
of $X_n$ corresponding to  eigenstate $|E_m\rangle$.  For a
specific $\Gamma_{m}$, $U_{m}(t)$ may be expressed in a factorized
form, which will be shown later through the spin-boson model.
Furthermore, we assume that the environment and the system are
initially independent, such that the total density operator
factorizes into a direct product,
\begin{eqnarray}
\rho(0)=\rho_S(0)\otimes\rho_B(0)=\nonumber\\
\sum_{mn}\rho_{mn}(0)|E_{m}\rangle\langle E_{n}|\otimes\rho_B(0).
\end{eqnarray}
At time $t$, the reduced density operator of the system is given
by
\begin{eqnarray}
\rho_S(t)&=&\mbox{Tr}_B
(U(t)\rho_S(0)\otimes\rho_B(0)U^{\dagger}(t))\nonumber\\
&\equiv&\sum_{mn}\rho_{mn}(0)e^{-i(E_{m}-E_{n})t}|E_{m}\rangle\langle
E_{n}|F_{mn}(t),\label{dmst}
\end{eqnarray}
where $F_{mn}(t)$ is defined as $
\mbox{Tr}_B[U_{m}(t)\rho_B(0)U_{n}^{\dagger}(t)].$ Eq.(\ref{dmst})
shows that the diagonal elements of the reduced density matrix
$\rho_{mm}$ are time-independent, while the off-diagonal elements
evolve with time involving contributions from the
environment-system couplings, it at most cases would lead to
decaying in the off-diagonal elements, and eventually results in
vanishing of these matrix elements. Now we turn to study the
geometric phase of  the open system. For open systems, the states
in general  are not pure and the evolution of the system   is not
unitary. For non-unitary evolutions, the geometric phase can be
calculated as follows. First, solve the eigenvalue problem for the
reduced density matrix $\rho(t)$ and obtain its eigenvalues
$\varepsilon_k(t)$ as well as the corresponding eigenvectors
$|\psi_k(t)\rangle;$ secondly, substitute $\varepsilon_k(t)$ and
$|\psi_k(t)\rangle$ into
\begin{widetext}
\begin{equation}
\Phi_g=\mbox{Re} \{
\mbox{arg}(\sum_k\sqrt{\varepsilon_k(0)\varepsilon_k(T)}\langle\psi_k(0)|\psi_k(T)\rangle
e^{-\int_0^T\langle\psi_k(t)|\partial/\partial t|\psi_k(t)\rangle
dt})\}.\label{gp5}
\end{equation}
\end{widetext}
Here, $\Phi_g $ is the geometric phase for the system undergoing
non-unitary evolution \cite{tong04}, $T$ denotes a time after that
the system completes  a cyclic evolution when it is isolated from
any environments. Taking the effect of environment into account,
the system no longer undergoes a cyclic evolution, but we still
keep the $T$ as the period for calculating the geometric phase in
this paper.  The geometric phase Eq. (\ref{gp5}) is gauge
invariant and can be reduced to the well-known results in the
unitary evolution, thus it is experimentally testable. The
geometric phase factor defined by Eq.(\ref{gp5}) may be understood
as a weighted sum over the phase factors pertaining to the
eigenstates of the reduced density matrix, thus the detail of
analytical expression for the geometric phase would depend on the
digitalization of the reduced density matrix Eq.(\ref{dmst}).

\section{geometric phase in dephasing system: example}
To be specific, we now present a detailed model to illustrate the
  idea in Sec. {\rm II}. The system under consideration consists of a two-level
system coupling to its environment with coupling constants
$\{g_i\}$, the Hamiltonian which governs the evolution of such a
system may be expressed as
\begin{eqnarray}
H&=&\frac{\omega}{2}(|e\rangle\langle e|-|g\rangle\langle g|)
+\frac 1 2 (|e\rangle\langle e|-|g\rangle\langle g|)\sum_i
g_i(a_i^{\dagger}+a_i)\nonumber\\
&+&\sum_i\omega_i a_i^{\dagger}a_i,
\end{eqnarray}
where $a_i^{\dagger}$, $a_i$ are the creation and annihilation
operators of the environment bosons, and $|e\rangle$, $|g\rangle$
denote the excited and ground states of the two-level system with
Rabi frequency $\omega$. This Hamiltonian corresponds to
$X_m=X=\frac 1 2 (|e\rangle\langle e|-|g\rangle\langle g|),$ and
$\Gamma_m=\Gamma=\sum_i g_ia_i$ in the general model
Eq.(\ref{gm}). Generally speaking, the choice of the coupling
between the system and the environment determines the effect of
the environment. For example, the choice of the system operator
$X_m$ that do not change the good quantum number of $H_S$ when
they operate on the eigenstates of $H_S$ would result in dephasing
of the system, but not energy relaxations. The system-environment
coupling taken in this section is exactly of this kind.

By the procedure presented above, the reduced density matrix in
basis $\{|e\rangle, |g\rangle\}$ for the open system follows \cite
{sun95},
\begin{equation}
\rho_S=\left( \matrix{ \cos^2\frac{\theta}{2} &  \frac 1 2
\sin\theta F_{12}(t) \cr
  \frac 1 2\sin\theta F_{21}(t) & \sin^2\frac{\theta}{2} \cr } \right),\label{ha2}
\end{equation}
where an initial state of
$(\cos\frac{\theta}{2}|e\rangle+\sin\frac{\theta}{2}|g\rangle)\otimes|0\rangle_B$
for the total system was assumed in the calculation, and
\begin{equation}
F_{12}(t)=F_{21}(t)=F(t)=e^{-i\omega t} e^{-\sum_j \eta_j(t)},
\end{equation}
with  $\eta_j(t)=4|\frac{g_j}{\omega_j}|^2(1-\cos\omega_jt)$, and
 $|0\rangle_B$ denoting the vacuum state of the environment. Some
remarks on the reduced density matrix are now in order. For any
$j$, $\eta_j(t)\geq 0$, so as $t$ tends to infinity (with respect
to the system's coherence time), $F(t)$ tends to zero, this
indicates that the off-diagonal elements would vanish on a long
time scale with respect to the decoherence time, and hence the
open system would not acquire geometric phase when time is longer
than the decoherence time. This is different form the results
concluded in the previous work, where the subsystem may acquire
geometric phase even for the whole system in its pointer states
\cite{yi05}. To calculate the geometric phase pertaining to
Eq.(\ref{ha2}), we first write down the eigenstate and its
corresponding eigenvalue for the reduced density matrix $\rho_S$
as,
\begin{eqnarray}
|\varepsilon_{\pm}(t)\rangle &=&
\cos\Theta_{\pm}(t)|e\rangle+\sin\Theta_{\pm}(t)|g\rangle,\nonumber\\
\varepsilon_{\pm}(t)&=&\frac 1 2
(1\pm\sqrt{\cos^2\theta+\sin^2\theta|F(t)|^2}), \label{ev}
\end{eqnarray}
with
\begin{equation}
\cos\Theta_{\pm}(t)=\frac{\sin\theta F(t)}{\sqrt{\sin^2\theta
|F(t)|^2+4(\varepsilon_{\pm}(t)-\cos^2\frac{\theta}{2})^2}}.
\label{ef}
\end{equation}
Clearly, for a closed system, namely $g_i=0$, $F(t)=e^{-i\omega
t}$, the eigenvalue and corresponding eigenstates reduce to
$\varepsilon_{\pm}(t)=1,0$, and
$\cos\Theta_+(t)=\cos\frac{\theta}{2}e^{-i\omega t}$,
$\sin\Theta_+(t)=\sin\frac{\theta}{2},$
$\cos\Theta_-(t)=-\sin\frac{\theta}{2},$
$\sin\Theta_-(t)=\cos\frac{\theta}{2}e^{i\omega t}.$ All these
together yield the well-known geometric phase
$\Phi_g^{(0)}=(1+\cos\theta).$ Eq.(\ref{ev}) and Eq.(\ref{ef}) are
the exact  results for the open two-level system, the geometric
phase would depend on how $F(t)$ varies with time. For a
continuous spectrum of  environmental modes with constant spectral
density $\sigma(\omega)=\epsilon,$ $F(t)=e^{-i\omega t}e^{-\gamma
t}$ with $\gamma=2\pi\epsilon |g|^2,$ where $g_i=g$ was assumed.
Up to the first order in $\gamma$, the geometric phase at time
$T=2\pi/\omega$ is given by
\begin{equation}
\Phi_g^{(1)}=\pi(1+\cos\theta)-\frac{\gamma}{\omega}\pi^2\sin^2\theta.
\end{equation}
This result can be easily understood as follows. The geometric
phase factor for mixed states is defined as a weighted sum over
the phase factors pertaining to the eigenstates of the reduced
density matrix, the dephasing that leads to decaying in the
off-diagonal elements would change the phase factors acquired by
each eigenstate of the reduced density matrix, thus it modifies
the geometric phase. This is different from the definition in the
quantum jump approach \cite{carollo03}, in which the problem of
defining Berry's phase for mixed states was avoided by approaching
the dynamics of  open system from a sequence of pure states, this
leads to the result that the geometric phase is unaffected by
dephasing, but it lowers the observed visibility in any
interference measurements. From the aspect of mixed state, the
evolution of the system is among several trajectories with
corresponding probabilities, so the geometric phase is defined as
a weighted sum over the trajectories that the system undergoes.

Now we are in a position to discuss the geometric phase acquired
at time $T=2\pi/\omega$ by the two-level system. Substituting
Eqs(\ref{ev}) and (\ref{ef}) in to Eq.(\ref{gp5}), we obtain
\begin{equation}
\Phi_g=\int_0^T\cos^2 \Theta_+(t) dt. \label{bpf}
\end{equation}
It is worthy to notice that   Eq.(\ref{bpf}) is the geometric
phase  beyond the quantum Markov approximation and the weak
coupling limit. In this sense it provides us more insight into the
geometric phase for dephasing system. The numerical results for
Eq. (\ref{bpf})  are presented in figure 1, where   the dependence
of the geometric phase is illustrated as a function of the
azimuthal angle $\theta$ and the damping rate $\gamma$. The
spectrum of environmental modes was taken to be
$\sigma(\omega)=\varepsilon$ in this plot. Clearly, the two-level
system acquires zero geometric phase with $\gamma$ tends to
infinity, this indicates that the two-level system acquires no
geometric phase after the decoherence time. The representation in
this paper may be used to analyze and estimate the error in the
holonomic quantum computation due to decoherence \cite{fuentes05},
in which the key error occurs within the degenerate subspace. The
Hamiltonian that describes such a system reads $H=
\sum_{i,j,\alpha}
g_{i,j}^{\alpha}(a_{\alpha}^{\dagger}+a_{\alpha})|i\rangle\langle
j| +\sum_i\omega_i a_i^{\dagger}a_i$, where the degenerate energy
was assumed to be zero, $\{|i\rangle, i=1,...,N\}$ denote the
degenerate levels coupled to the environment with coupling
constants $\{g_{i,j}^{\alpha}\}$. The Hamiltonian can be rewritten
as $H= \sum_{\beta,\alpha}
g_{\beta}^{\alpha}(a_{\alpha}^{\dagger}+a_{\alpha})|\beta
\rangle\langle \beta| +\sum_i\omega_i a_i^{\dagger}a_i$, with an
appropriate choice of $|\beta\rangle=\sum_{i=1}^N c_i|i\rangle$.
This exactly is the case discussed in sections {\rm II} and {\rm
III}.
\begin{figure}
\includegraphics*[width=0.8\columnwidth,
height=0.6\columnwidth]{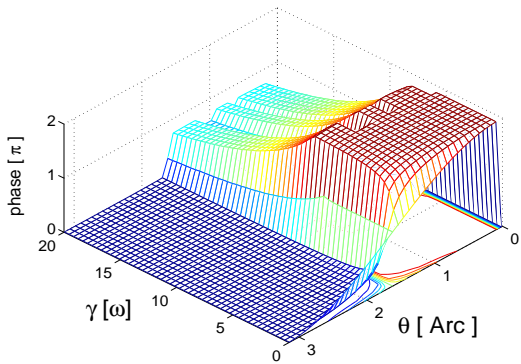} \caption{An illustration of
the geometric phases of the two-level system coupling to the
environment at time $T=2\pi/\omega.$}  \label{fig1}
\end{figure}

The open system effect in geometric phase may be observed with a
combination of the engineering reservoir technique \cite{myatt00}
and  the Mach-Zehnder atom interferometer
\cite{featonby98,webb99}, in which each of the arms consists of an
atom in a dark state. The dark state can be realized in the
atom-light system that consists of cesium atoms interacting with
light resonant with the $F=3 \rightarrow F^{'}=3$ transitions of
the $D1\ \ 6^2s_{1/2} \rightarrow 6^2 p_{1/2}$ line. It makes the
dynamical phase negligible with respect to the geometric phase. A
dephasing engineering reservoir in one arm of the interferometer,
which may be simulated by variations of the light fields,  would
give a relative phase to the atom passing though the arm, the
output interference pattern yields the geometric phase of the atom
system.

One of the key assumptions of our representation is the dephasing
condition, i.e., $[H_S, H_I]=0$. Using ground states as the qubits
can make the formulation  exact.  Suppose now there is an small
additional term in $H_I$,
$H_I^{'}=\sum_mY_m(\Gamma_m+\Gamma_m^{\dagger})$, that breaks  the
dephasing condition. Simple algebra shows that the transition
probability between $|E_m\rangle$ and $|E_n\rangle$ due to
coupling $Y_p (\Gamma_p+\Gamma_p^{\dagger})$ is proportional to
$|\gamma_p\langle E_n|Y_p|E_m\rangle|^2/|E_m-E_n|^2$, where
$\gamma_p$ denotes the maximum of average values of
$(\Gamma_p+\Gamma_p^{\dagger})$. In the case of $|\gamma_p\langle
E_n|Y_p|E_m\rangle|<<|E_m-E_n|$, the open system may be treated as
a dephasing system, because the transition between any different
eigenstates of the system  may be ignored. The case where this
transition could not be ignored will be discussed in the next
section.

\section{geometric phase in dissipative systems: exactly solvable model}
In this section, we will consider a spin-$\frac 1 2 $ particle
interacting with an environment formed by $N$ independent spins
through the Hamiltonian
\begin{equation}
H=\Delta\sigma_x+\frac 1 2 \sum_{k=1}^N g_k\sigma_z\sigma_z^{(k)},
\label{hd}
\end{equation}
where $\sigma_i^{(k)}$ and $\sigma_i$, $i=x,y,z$ denote Pauli
operators for the environment and spin-$\frac 1 2 $ particle,
respectively. $g_k, (k=1,2,...,N)$ are coupling constants, term
with $\Delta$ stands for the self-Hamiltonian of the particle.
This model is interesting because the pointer states do not
coincide with the eigenstates of the interaction Hamiltonian, but
can range from coherent sates to eigenstates of the system's
Hamiltonian determining by the interplay between the
self-Hamiltonian and the interaction with the environment. We will
calculate the geometric phase gained by the particle beyond the
Markov approximation and the weak coupling limit. The dynamics
govern by Hamiltonian Eq.(\ref{hd}) can be solved exactly by a
standard procedure\cite{cucchietti05}, it yields the reduced
density matrix of the particle as
\begin{equation}
\rho(t)=(I+\vec{p}(t)\cdot \vec{\sigma})/2,
\end{equation}
where $\vec{p}(t)$ is the polarization vector given by
$\vec{p}(t)=\int \vec{p}(t,B) \eta(B) dB$ with
$\Omega_B^2=\Delta^2+B^2$ and
\begin{eqnarray}
\eta(B)&=&\frac{1}{\sqrt{2\pi s_N^2}}e^{-B^2/2s_N^2},\nonumber\\
p_x(t,B)&=&p_x(0) \frac{\Delta^2+B^2 \cos (2 \Omega_B
t)}{\Omega_B^2} - p_y(0) \frac{B}{\Omega_B} \sin(2 \Omega_B
t)\nonumber\\
&+& p_z(0) \frac{2 \Delta B }{\Omega_B^2}\sin^2(\Omega_B
t),\nonumber\\
p_y(t,B) &=& p_y(0) \cos(2 \Omega_B t) + \frac{\sin(2 \Omega_B
t)}{\Omega_B} \left[ p_x(0) B - \Delta p_z(0) \right],\nonumber\\
 p_z(t,B) &=& p_z(0) \frac{B^2+\Delta^2 \cos
(2 \Omega_B t)}{\Omega_B^2} \nonumber\\
&+& p_x(0) \frac{2 \Delta B }{\Omega_B^2}\sin^2(\Omega_B t) +
p_y(0) \frac{\Delta}{\Omega_B} \sin(2 \Omega_B t).\label{pt}
\end{eqnarray}
To get this result, it is only required that the couplings $g_k$
of Eq.(\ref{hd}) are sufficiently concentrated near their average
value so that their standard deviation exists and is finite.

 By rewriting the reduced density matrix $\rho(t)$ in the
form
\begin{equation}
\rho(t)=\lambda_1(t)|\psi_1(t)\rangle\langle
\psi_1(t)|+\lambda_2(t)|\psi_2(t)\rangle\langle \psi_2(t)|,
\end{equation}
we get the geometric phase\cite{tong04} of the particle acquired
at time $\tau$,
\begin{widetext}
\begin{equation}
\Phi_g^{\prime}(\tau)=arg\left( \sum_{i=1,2}
\sqrt{\lambda_i(0)\lambda_i(\tau)}\langle
\psi_i(0)|\psi_i(\tau)\rangle
e^{-\int_0^{\tau}\langle\psi_i(t)|\frac{\partial}{\partial
t}|\psi_i(t)\rangle dt} \right).
\end{equation}
\end{widetext}
After simple manipulations, we arrive at
\begin{eqnarray}
\Phi_g^{\prime}&=&arg [
\Lambda_1(\tau)(\cos\frac{\theta(0)}{2}\cos\frac{\theta(\tau)}{2}
e^{-i(\phi(\tau)-\phi(0))} \nonumber\\
 &+&\sin\frac{\theta(0)}{2}\sin\frac{\theta(\tau)}{2})e^{iC}  \nonumber\\
&+&\Lambda_2(\tau)(\cos\frac{\theta(0)}{2}\cos\frac{\theta(\tau)}{2}
e^{i(\phi(\tau)-\phi(0))}  \nonumber\\
&+&\sin\frac{\theta(0)}{2}\sin\frac{\theta(\tau)}{2})e^{-iC}].\label{bpd}
\end{eqnarray}
Here,
$$\Lambda_{1,2}=\frac 1 2
\sqrt{(1\pm  |\vec{p}(0)|) (1\pm  |\vec{p}(t)|)},$$

$$\cos\theta(t)=\frac{p_z(t)}{\sqrt{p_x^2(t)+p_y^2(t)+p_z^2(t)}},$$
$$\tan\phi(t)=\frac{p_y(t)}{p_x(t)},$$
$$C=\frac i 2\left( \int_0^{\tau}\frac{\partial \phi}{\partial t}
dt +\int_0^{\tau} \cos\theta \frac{\partial \phi}{\partial
t}dt\right).$$ The dependence of the geometric phase
$\Phi_g^{\prime}$ on the variance $s_N$ and system free energy
$\Delta$ is complicated, we discuss here on two limiting cases
$s_N\gg \Delta$ and $s_N \ll \Delta$ with a specific  initial
state $p_x(0)=1, p_y(0)=p_z(0)=0$. In $s_N \gg \Delta$ limit, the
dynamics of the spin-$\frac 1 2 $ particle is so slow that its
behavior should approach $p_x(t)=e^{-2t^2s_N^2}$ and
$p_y(t)=p_z(t)=0,$ which yields $\Phi_g^{\prime}=0$ because
$\phi=0$ in this limit with the initial state.  In $s_N\ll \Delta$
limit, Eq.(\ref{pt}) follows that,
\begin{eqnarray}
p_x(t)&\simeq& p_x(0) \left[ \gamma \left( \frac{\Delta}{\sqrt{2}
s_N}\right) + \frac{\cos \left( 2 \Delta t + \frac{3 \pi}{4}
\right)}{ \sqrt{8 \Delta s_N^2 t^3}}  \right],
\nonumber\\
p_y(t)&=&p_z(t)=0,
\end{eqnarray}
where $\gamma (x) = \sqrt{\pi} x e^{x^2} \left(1-{\rm Erf}(x)
\right),$  $Erf(x)$ is the error function, and $\gamma
\left(\frac{\Delta}{ \sqrt{2} s_N}\right) \ll 1$ in this limit. By
Eq.(\ref{bpd}), it is clear that $\Phi_g^{\prime}=0$ in this
limit, too.  The numerical result for the geometric phase as a
function of $\Delta$ and $s_N$ was shown in figure 2. To plot this
figure, we assume that the system has evolved for time
$\tau=2\pi/\Delta$, which is the characterized time for the system
undergoing a free evolution. Figure 2 shows that the geometric
phase is zero in the two limiting cases $\Delta\gg s_N$ and
$\Delta \ll s_N$ as expected. There are sharp changes among the
line $\Delta \sim s_N$, indicating a crossover from the limit
$\Delta \ll s_N$ to $\Delta \gg s_N.$
\begin{figure}
\includegraphics*[width=0.8\columnwidth,
height=0.6\columnwidth]{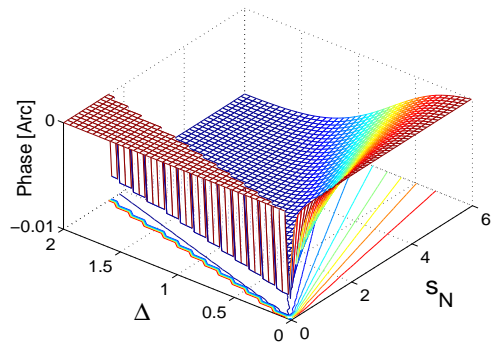} \caption{ The geometric phase
of the  spin-$\frac 1 2 $ particle versus $s_N$ and $\Delta$. The
spin was dissipatively coupled  to an environment and the phase
was calculated at time $\tau=2\pi/\Delta$, depending on the time
scale of free evolution of the particle. } \label{fig2}
\end{figure}

\section{summary and discussion}

 We have presented a general calculation the geometric phase in open
systems subject to dephasing and dissipation, the calculations are
beyond the quantum Markov approximation and the weak coupling
limit. For the dephasing system, it acquires no geometric phase
with the decoherence rate $\gamma\rightarrow \infty$, this can be
explained as an effect of decoherence on the geometric phase,
i.e., the quantum system could not maintain its phase information
after the decoherence time. There is a sharp change along the line
$\theta=\pi/2$ as figure 1 shown, this can be understood in terms
of the Bloch sphere that represents the state of the system. The
geometric phase increases due to decoherence when   initial states
fall onto the upper semi-sphere, but it decrease when the initial
states on  the lower semi-sphere. These results are similar to the
prediction given by the quantum trajectory analysis for
dissipative systems. The geometric phase $\Phi_g^{\prime}$ in
dissipative systems is always zero as long as
$p_x(t)/p_y(t)=$constant, this is exactly the case when
$\Delta/s_N\rightarrow \infty$ or $\Delta/s_N\rightarrow 0$.
$\Delta/s_N\rightarrow \infty$ implies that the self energy
$\Delta$ of the particle is much larger than the cumulative
variance $s_N$ of the coupling constants $g_k$. For $g_k$ taking
the value $+g$ or $-g$ ($g$ arbitrary) with equal probability,
$s_N^2=\sum_k g_k^2.$ This tells us that the geometric phase  is
zero when the self-Hamiltonian dominates. On the other hand, when
$\Delta/s_N\rightarrow 0$ the interaction Hamiltonian dominates,
pointer states in this situation coincide with the eigenstates of
the interaction Hamiltonian, thus the spin-$\frac 1 2 $ particle
could not acquire geometric phase. In the crossover regime $\Delta
\sim s_N$, the geometric phase change sharply due to the interplay
between the self-Hamiltonian and the interaction with the
environment.

 These results constitute the
basis of a framework to analyze errors in the holonomic quantum
computation, where two kinds of errors are believed to affect its
performance. This first error would take the system out of the
degenerate computation subspace, while the second takes place
within the subspace. The first kind of error can be eliminated by
working in the ground states and having a system where the energy
gap with the first excited state is very large. The second kind of
error falls to the regime analyzed in Sec. {\rm II} and {\rm III},
since there is no dissipation but dephasing in the system, while
the first belongs to the regime discussed in Sec. {\rm IV}. The
calculation presented here in principle allows one to study the
geometric phase at any timescale, and hence it has advantages with
respect to any treatment with approximations in most literatures.

\ \ \\
X.X.Y. acknowledges simulating discussions with Dr. Robert
Whitney. This work was supported by EYTP of M.O.E,  NSF of China
(10305002 and 60578014), and the NUS Research Grant No.
R-144-000-071-305.\\


\begin{references}
\bibitem{berry84} M. V. Berry, Proc. R. Soc. London A {\bf 392},
45(1984).

\bibitem{shapere89} Geometric phase in physics, Edited by A. Shapere
and F. Wilczek ( World Scientific, Singapore, 1989).

\bibitem{uhlmann86} A. Uhlmann,
Rep. Math. Phys. {\bf 24}, 229 (1986).
\bibitem{sjoqvist00a} E. Sj\"{o}qvist, A.K. Pati, A. Ekert,
J.S. Anandan, M. Ericsson, D.K.L. Oi, and V. Vedral, Phys. Rev.
Lett. {\bf 85}, 2845 (2000).

\bibitem{fuentes02} I. Fuentes-Guridi, A. Carollo, S. Bose, and V. Vedral,
Phys. Rev. Lett. {\bf 89}, 220404 (2002); A. Carollo, I.
Fuentes-Guridi, M. Franca Santos and  V. Vedral, Phys. Rev. A {\bf
67}, 063804(2003).

\bibitem{yi04}X.X. Yi, L.C. Wang, and T.Y. Zheng,
Phys. Rev. Lett. {\bf 92}, 150406 (2004); X. X. Yi, and E.
Sj\"oqvist, Phys. Rev. A {\bf 70}, 042104 (2004);  L. C. Wang, H.
T. Cui, and X. X. Yi,  Phys. Rev. A {\bf 70}, 052106 (2004).

\bibitem{zanardi99} P. Zanardi and M. Rasetti,
Phys. Lett. A {\bf 264}, 94 (1999).

\bibitem{jones99} J. A. Jones, V. Vedral, A. Ekert,
and G. Castagnoli, Nature (London) {\bf 403},
869 (1999).

\bibitem{ekert00} A. Ekert, M. Ericsson, P. Hayden, H. Inamori, J.A. Jones,
D.K.L. Oi, and V. Vedral, J. Mod. Opt. {\bf 47}, 2051 (2000).

\bibitem{falci00} G.Falci, R. Fazio, G.M. Palma, J. Siewert, and V. Vedral,
Nature (London) {\bf 407}, 355 (2000).

\bibitem{garrison88} J. C. Garrison and E. M. Wright, Phys. Lett.
A {\bf 128}, 177(1988).

\bibitem{fonsera02} K. M. Fonseca Romero, A. C. Aguira Pinto, and
M. T. Thomaz, Physica A {\bf 307}, 142(2002).

\bibitem{aguiera03} A. C. Aguira Pinto and M. T. Thomaz, J. Phys.
A: Math. Gen. {\bf 36}, 7461(2003).

\bibitem{ellinas89} D. Ellinas, S. M. Barnett, and M. A.
Dupertuis, Phys. Rev. A {\bf 39}, 3228(1989).

\bibitem{gamliel89} D. Gamliel and J. H. Freed, Phys. Rev. A {\bf
39}, 3238(1989).

\bibitem{kamleitner04} I. Kamleitner, J. D. Cresser, and B. C.
Sanders, Phys. Rev. A {\bf 70}, 044103(2004).

\bibitem{nazir02} A. Nazir, T. P. Spiller, W. J. Munro, Phys. Rev.
A {\bf 65}, 042303(2002).

\bibitem{carollo03} A. Carollo, I. Fuentes-Guridi,
M. Franca Santos, and V. Vedral, Phys. Rev. Lett. {\bf
90},160402(2003); {\it ibid} {\bf 92}, 020402(2004).

\bibitem{marzlin04} K. P. Marzlin, S. Ghose, and B. C. Sanders,
Phys. Rev. Lett. {\bf 93}, 260402 (2004).

\bibitem{whitney03} R. S. Whitney, and Y. Gefen, Phys. Rev. Lett.
{\bf 90}, 190402(2003); R. S. Whitney, Y. Makhlin, A. Shnirman,
and Y. Gefen, e-print:cond-mat/0405267.

\bibitem{gaitan98} F. Gaitan, Phys. Rev. A {\bf 58}, 1665(1998).


\bibitem{yi05} X. X. Yi, L. C. Wang, and W. Wang, Phys. Rev. A {\bf 71}, 044101 (2005).

\bibitem{tong04} D. M. Tong, E. Sj\"oqvist, L. C. Kwek, C. H. Oh,
Phys. Rev. Lett. {\bf 93}, 080405 (2004).

\bibitem{sun95} C. P. Sun, X. X. Yi, X. J. Liu, Fortschritte Der
Physik {\bf 43}, 585 (1995).

\bibitem{note1} $\{|E_m\rangle\}$ was assumed nondegenerate, for
the degenerate case, we may choose some linear combination of the
degenerate eigenstates as the new eigenstates of $H_S$, such that
$\{|E_m\rangle\}$ are also belong to $X_m$.

\bibitem{fuentes05} I. Fuentes-Guride, F. Girelli, and E. Livine,
Phys. Rev. Lett. {\bf 94}, 020503(2005).


\bibitem{myatt00} C. J. Myatt, B. E. King, Q. A. Turchette, C. A.
Sackett, D. Kielpinski, W. M. Itano, C. Monore, and D. J.
Wineland, Nature(London) {\bf 403}, 269(2000).

\bibitem{featonby98} P. D. Featonby {\it et al.}, Phys. Rev. Lett.
{\bf 81}, 495(1998).

\bibitem{webb99} C. L. Webb {\it et al.}, Phys. Rev. A {\bf 60},
R1783 (1999).

\bibitem{cucchietti05} F. M. Cucchietti, J. P. Paz, and W. H.
Zurek, Phys. Rev. A {\bf 72}, 052113(2005).


\end{references}
\end{document}